\documentclass[twocolumn]{aastex61}
\pdfoutput=1 
\usepackage{amsmath,amstext}
\usepackage[T1]{fontenc}
\usepackage{apjfonts} 
\usepackage[figure,figure*]{hypcap}

\shorttitle{Resonant Excitation of Highly Charged Iron Ions}
\shortauthors{Tsuda et al.}

\begin{document}

\title{Resonant Electron Impact Excitation of $3\MakeLowercase{d}$ levels in F\MakeLowercase{e}$^{14+}$ and F\MakeLowercase{e}$^{15+}$}

\author{Takashi Tsuda}
\affiliation{Institute for Laser Science, The University of Electro-Communications, Tokyo 182-8585, Japan}

\author{Erina Shimizu}
\affiliation{Institute for Laser Science, The University of Electro-Communications, Tokyo 182-8585, Japan}

\author{Safdar Ali}
\affiliation{Institute for Laser Science, The University of Electro-Communications, Tokyo 182-8585, Japan}

\author{Hiroyuki A. Sakaue}
\affiliation{National Institute for Fusion Science, Gifu 509-5292, Japan}

\author{Daiji Kato}
\affiliation{National Institute for Fusion Science, Gifu 509-5292, Japan}
\affiliation{Department of Fusion Science, SOKENDAI, Gifu 509-5292, Japan}
\affiliation{Department of Advanced Energy Engineering Science, Kyushu University, Fukuoka 816-8580, Japan}

\author{Izumi Murakami}
\affiliation{National Institute for Fusion Science, Gifu 509-5292, Japan}
\affiliation{Department of Fusion Science, SOKENDAI, Gifu 509-5292, Japan}

\author{Hirohisa Hara}
\affiliation{National Astronomical Observatory of Japan, Tokyo 181-8588, Japan}
\affiliation{Department of Astronomical Science, SOKENDAI, Tokyo 181-8588, Japan}

\author{Tetsuya Watanabe}
\affiliation{National Astronomical Observatory of Japan, Tokyo 181-8588, Japan}
\affiliation{Department of Astronomical Science, SOKENDAI, Tokyo 181-8588, Japan}

\author{Nobuyuki Nakamura}%
\affiliation{Institute for Laser Science, The University of Electro-Communications, Tokyo 182-8585, Japan}


\begin{abstract}
We present laboratory spectra of the $3p$--$3d$ transitions in Fe$^{14+}$ and Fe$^{15+}$ excited with a mono-energetic electron beam.
In the energy dependent spectra obtained by sweeping the electron energy, resonant excitation is confirmed as an intensity enhancement at specific electron energies.
The experimental results are compared with theoretical cross sections calculated based on fully relativistic wave functions and the distorted-wave approximation.
Comparisons between the experimental and theoretical results show good agreement for the resonance strength.
A significant discrepancy is, however, found for the non-resonant cross section in Fe$^{14+}$.
This discrepancy is considered to be the fundamental cause of the previously reported inconsistency of the model with the observed intensity ratio between the $^3\!P_2$ -- $^3\!D_3$ and $^1\!P_1$ -- $^1\!D_2$ transitions.

\end{abstract}

\keywords{atomic processes --- Sun: UV radiation --- Sun: corona --- solar corona --- line: formation}

\section{Introduction}

Resonant processes in collisions of highly charged iron ions with electrons are important for astrophysical hot plasmas, such as the solar corona.
Resonant processes are initiated by dielectronic capture, in which the incident electron is captured to the ion while exciting an inner shell electron.
The decay of the resultant inner-shell excited state determines the process as follows. 
When two electrons are emitted by double autoionizing decay, the process is resonant ionization called resonant excitation double autoionization (REDA) \citep{Chen9,Linkemann1,Kwon1}.
When no electron is emitted and only photons are emitted, the process is resonant recombination called dielectronic recombination (DR) \citep{Schippers2,Lukic1,Savin2}.
Since REDA and DR change the charge state of ions, they strongly affect the charge state distribution of iron ions in the astrophysical plasmas.
For example, the importance of DR in the x-ray spectra obtained with observatories such as {\it Chandra} and {\it XMM-Newton} is discussed by \citet{Schippers2}, \citet{Savin2}, and \citet{Lukic1}.
On the other hand in the process of the present interest, only one electron is emitted through autoionization in the decay of the inner-shell excited state.
Thus the charge state of the ion is not changed before and after the collision, but the ion can be in an excited state after the autoionization.
Such a process is called resonant excitation (RE) and strongly affects the line emission intensity in the astrophysical plasmas as discussed by \citet{Goldstein1}, and \citet{Smith1}.

Due to their importance, theoretical studies of these resonant processes have been extensively made to date, and laboratory measurements have provided the benchmark for the theoretical calculations.
A heavy ion storage ring \citep{Bisoffi1,Nortershauser1} and an electron beam ion trap (EBIT) \citep{Marrs1,yebisu_nakamura} are powerful laboratory devices for studying resonant processes of highly charged ions.
In the experiments with a storage ring, charge changed particles are detected; thus the charge-changing processes, such as REDA and DR, have been extensively studied \citep{Chen9,Schippers2}.
On the other hand in the experiments with an EBIT, emission from the trapped ions excited by an electron beam can be detected; thus, it is possible to study not only charge-changing processes but also non-charge-changing processes, such as RE.
However, in fact only a few RE measurements have been performed so far \citep{Beiersdorfer4,Takacs1} although a lot of DR measurements have been extensively performed \citep{Beiersdorfer24,Watanabe4}.
Probably it is because the atomic number and the charge state of interest are generally so high that DR is the dominant channel due to the large fluorescence yield of the intermediate inner shell excited states.
In order to study RE for relatively low charged light ions, the electron energy for producing the ions and probing the resonance should be as low as 1~keV or lower.
However, such low energy operation has hardly been applied for resonance measurements because ordinary EBITs are suitable for the operation with an electron energy of few to several hundreds of keV.

In this paper, we present RE measurements for Fe$^{14+}$ and Fe$^{15+}$ with a compact EBIT, which is suited for low energy operation, in the electron energy range of 350 - 500~eV:
\begin{eqnarray*}
e + {\rm Fe}^{q+} (2p^6 3s^k) &\rightarrow& {\rm Fe}^{(q-1)+} (2p^5 3s^k 3l 3d)\\
&\rightarrow& {\rm Fe}^{q+} (2p^6 3s^{(k-1)} 3d) + e ,
\end{eqnarray*}
where $k=2$ for $q=14$ and $k=1$ for $q=15$.
The resultant $3d$ excited states mainly decay to $3p$ levels by emitting a photon in the extreme ultraviolet (EUV) range.
The measurements have been done by observing this $3p$--$3d$ transitions.
The $3p$--$3d$ transitions in iron ions are useful for the density and temperature diagnostics, and have been widely observed so far with several solar observatories, such as {\it Skylab} \citep{Dufton1}, the {\it Solar EUV Research Telescope and Spectrograph} ({\it SERTS}) \citep{Keenan3}, the Atmospheric Imaging Assembly (AIA) on {\it Solar Dynamic Observatory} ({\it SDO}) \citep{Lemen1} and the EUV Imaging Spectrometer (EIS) on {\it Hinode} \citep{Hara1}.
Thus, it is important for the solar corona diagnostics to know how the RE processes affect the line emission.
In our previous work \citep{Shimizu2,Nakamura21}, the electron density dependence of the $3p$--$3d$ transitions in Fe$^{14+}$ was studied, and a significant discrepancy was found between the experiment and a theoretical model for the line intensity ratio between the $3s3p$ $^3\!P_2$ -- $3s3d$ $^3\!D_3$ (234~\AA) and the $3s3p$ $^1\!P_1$ -- $3s3d$ $^1\!D_2$ (244~\AA) transitions.
Such discrepancy has also been often reported from observatory measurements.
For example, \citet{Dufton1} compared their model calculation with the {\it Skylab} observations of solar flares, and found that most of the observed values for the ratio $I(234~\mathrm{\mathring{A}})/I(244~\mathrm{\mathring{A}})$ exceed the theoretical high density limit.
They considered that the discrepancy was attributed to the accuracy of the theoretical values, and pointed out that their calculation based on the $LS$-coupling with the limited configurations should be improved.
\citet{Keenan2} improved the calculation by using fully relativistic electron excitation rates with extended configurations, but found that the results were similar to those of \citet{Dufton1}.
Thus, they considered that the discrepancy found by \citet{Dufton1} should be attributed to blending of other lines rather than the accuracy of the theoretical data.
\citet{Kastner1} pointed out that the active region observations by \citet{Dere2} are consistent with their model calculations, which is not much different from that of \citet{Dufton1}; thus, they also considered that the discrepancy found by \citet{Dufton1} should be attributed to blending, which could be enhanced in flares.
However, our recent laboratory observation with an EBIT showed that line blending is not responsible for the discrepancy \citep{Shimizu2}.
Relatively recent calculation by \citet{Landi6} also showed a theoretical high density limit which is similar to those obtained in the previous calculations; thus the discrepancy with the flare observations remains a still unsolved question.
One of the motivations of the present study is to examine how the RE processes affect this line intensity ratio.

\section{\label{sec:exp}Experiment}

The present experiment has been carried out with a compact EBIT, called CoBIT \citep{CoBIT,Tsuda1}.
It consists of an electron gun, a drift tube (DT), an electron collector, and a high-critical-temperature superconducting magnet.
The DT is composed of three successive cylindrical electrodes that act as an ion trap by applying a positive trapping potential (150~V in the present experiment) at both ends (DT1 and 3) with respect to the middle electrode (DT2).
The electron beam emitted from the electron gun is accelerated towards the DT while it is compressed by the axial magnetic field (0.08~T in the present experiment) produced by the magnet surrounding the DT.
The compressed high-density electron beam ionizes the ions trapped in the DT.
Highly charged iron ions were produced by successive ionization of iron injected as a vapor of ferrocene (Fe(C$_5$H$_5$)$_2$).

\begin{figure}
\includegraphics[width=0.45\textwidth]{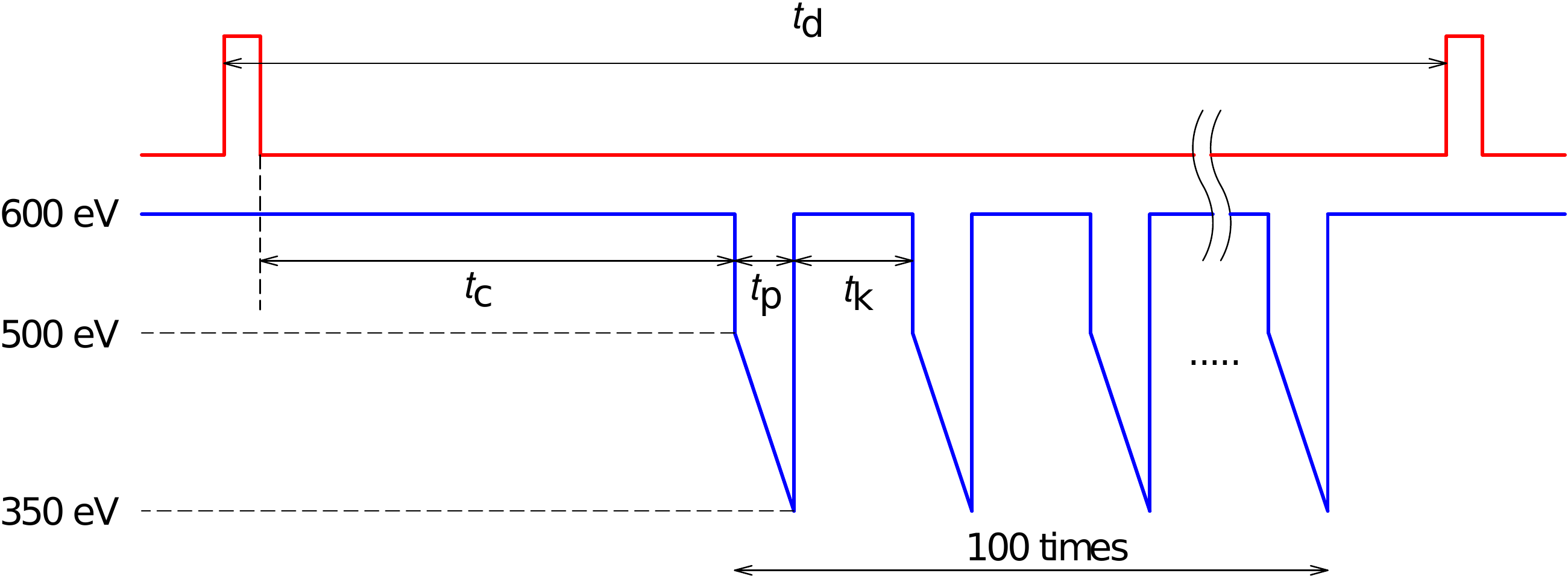}
\caption{\label{fig:seq}
Time sequence of the present experiment: (red) the pulse for dumping the trapped ions; (blue) electron energy.
$t_{\rm d}$, $t_{\rm b}$, $t_{\rm p}$, and $t_{\rm k}$ are dumping period (3600 ms), charge-breeding time (1600 ms), probing time (7 ms), and keeping time (13 ms), respectively.
}
\end{figure}

For observing resonant excitation, the electron beam energy was controlled by the time sequence shown in Fig.~\ref{fig:seq}.
The electron beam energy was first fixed at 600~eV for 1600~ms ($t_{\rm c}$) to maximize the abundance of Fe$^{14+}$ and Fe$^{15+}$, which are ions of present interest.
After this charge-breeding time, the electron energy was swept from 500 to 350~eV within about 7~ms (probing time, $t_{\rm p}$), and kept at 600~eV for about 13~ms (keeping time, $t_{\rm k}$) for preserving the charge distribution.
After the probing and keeping periods were repeated 100 times, the ions were dumped and the cycle was started again from the breeding time.
As the electron energy in the EBIT is essentially determined by the potential difference between the electron gun (cathode) and the middle electrode of the trap (DT2), the energy control was done by sweeping the DT2 potential from \mbox{-100}~V to \mbox{-250} during the probing time while keeping the cathode potential at \mbox{-600}~V throughout the measurement.
The actual electron energy interacting with the trapped ions is generally lower than the value determined from the potential difference between the cathode and DT2 mainly due to the space charge potential of the electron beam.
In the present study, the absolute electron energy scale was normalized to the theoretical resonance energy as it is generally difficult to know the space charge contribution precisely. 
The electron beam current was 12 mA throughout the measurement.

The EUV emission from the trapped iron ions was observed with a grazing incidence flat field grating spectrometer \citep{Ohashi4} employing a 1200~gr/mm concave grating with a 13450~mm radius of curvature (Hitachi 001-0660).
In the present setup, no entrance slit was used because the EBIT represents a line source which can be regarded as a slit.
The diffracted EUV photons were detected by a two-dimensional position sensitive detector (PSD) consisting of five micro channel plates (MCPs) and a resistive anode (Quantar Technology Inc., model 3391).
The front of the first MCP was coated by CsI for enhancing the sensitivity.
For each detected photon, a PC recorded list mode data consisting of the two-dimensional position ($X$ and $Y$) on the PSD, the pulse height of the signal, and the DT2 potential (corresponding to the electron energy) at the time when the photon was detected.
As the direction $X$ was used for the dispersion direction, the position $X$ was converted to the wavelength of the diffracted photon after correcting the arc-like curved image obtained in the two-dimensional image \citep{Nakamura8}.
The wavelength scale was calibrated with nine prominent lines of Fe$^{13+}$ to Fe$^{15+}$, and the uncertainty in the calibration was estimated to be 0.06 \AA.
The pulse height was used to remove the electric noise.
By analyzing the list mode data, the energy dependent spectra as shown in Fig.~\ref{fig:2d} in Sec.~\ref{sec:results} were obtained.

\section{\label{sec:cal}Calculations}

For the comparison with the experimental data, excitation cross sections were calculated with the Hebrew University Lawrence Livermore Atomic Code (HULLAC v9.601) \citep{HULLAC2}.
The resonance strength for the RE from the initial state $ \left | i \right >$ to the final state $\left | f \right >$ via the intermediate state $\left | d \right >$ can be expressed as
\begin{equation}
\frac{\pi^2 \hbar^3}{p_{\rm e}^2} \cdot \frac{g_d}{g_i} \cdot \frac{A_a(d\rightarrow i) A_a(d\rightarrow f)}{\sum_j A_r(d\rightarrow j)+ \sum_k A_a(d\rightarrow k)} ,
\end{equation}
where $p_{\rm e}$ is the momentum of the incident electron, $g$ the statistical weight, $A_a$ the autoionizing rate, and $A_r$ the radiative transition rate.
For calculating the RE of Fe$^{14+}$, the radiative and autoionizing rates of the intermediate Fe$^{13+}$ levels were calculated by taking the 28,045 energy levels arising from the configurations $2p^6 3l nl' n'l''$ ($3 \leq n, n'\leq 5$), $2p^5 3s^2 3l nl'$ ($3 \leq n\leq 5$), and $2p^5 3s 3l 3l' nl''$ ($3 \leq n\leq 5$) into account.
For the RE of Fe$^{15+}$, the intermediate Fe$^{14+}$ levels were evaluated by taking the 12,260 energy levels arising from the configurations $2p^6 nl n'l'$ ($3 \leq n, n'\leq 5$), and $2p^5 3l 3l' nl''$ ($3 \leq n\leq 5$) into account.

\begin{figure}
\includegraphics[width=0.45\textwidth]{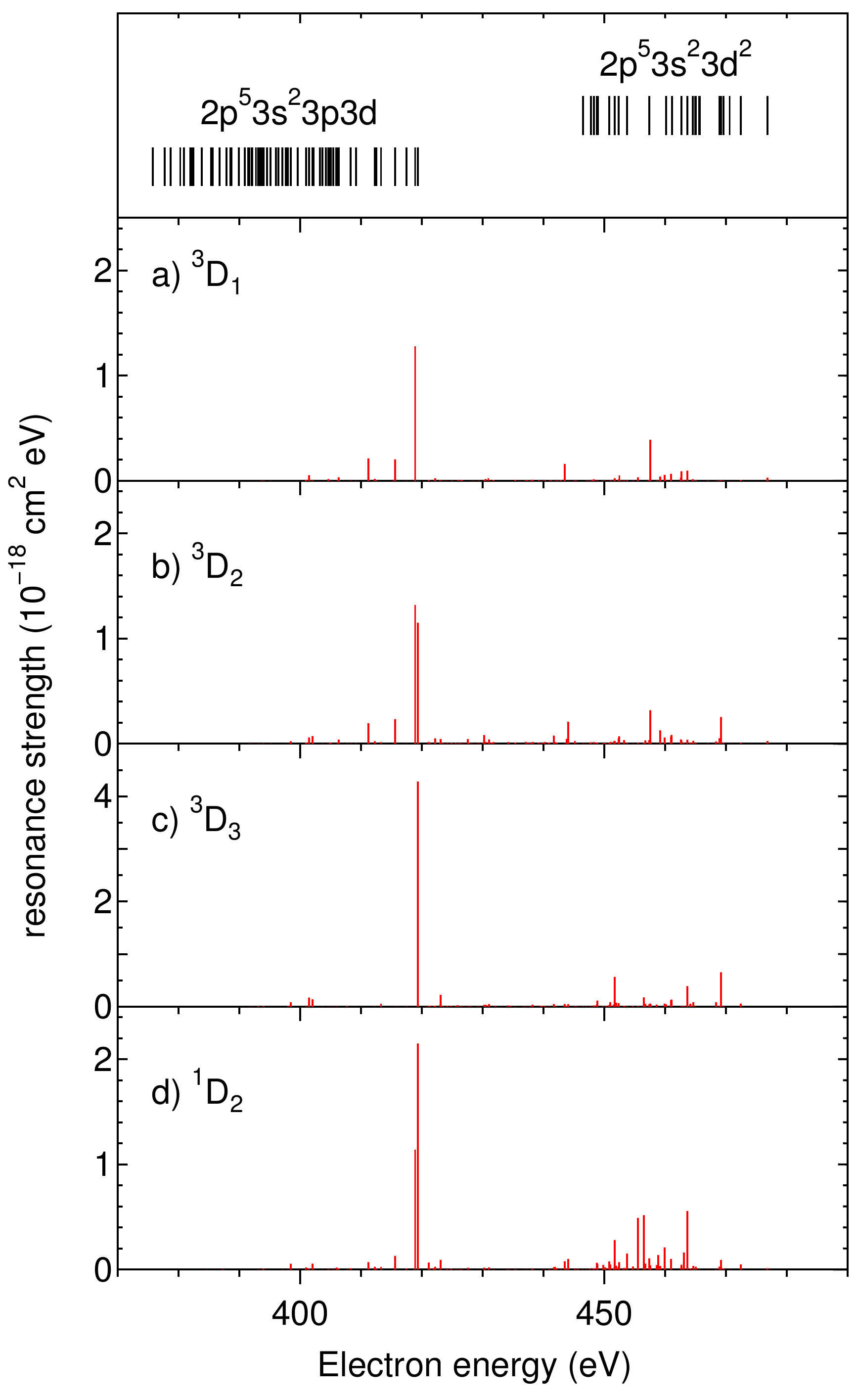}
\caption{\label{fig:rs}
Calculated resonance strength for the RE from the $3s^2$ ground state to the four $3s3d$ levels of magnesiumlike Fe$^{14+}$.
Note that the vertical axis range of c) is different from that of the others.
The top panel is the energy levels of the $2p^53s^23p3d$ and $2p^53s^23d^2$ intermediate states.
}
\end{figure}

The calculated results of the resonance strength for the RE from the $3s^2$ ground state to the $3s3d$ levels in Fe$^{14+}$ are shown in Fig.~\ref{fig:rs}.
In the energy region of 370 to 490~eV, RE via $2p^53s^23p3d$ and $2p^53s^23d^2$ is possible.
The most prominent peaks at around 420~eV correspond to resonances via $[2p_{1/2}^{-1}3s^23p_{1/2}3d_{j}]_{J=j}$ levels ($j=\frac{3}{2}$, $\frac{5}{2}$), whose energy splitting is 0.5 eV.
These two levels have a large autoionizing rate to the $3s^2$ ground state ($A_a(d\rightarrow i)$), resulting in a large resonance strength.
Assuming that the total angular momentum $j$ of the $3d$ electron in the intermediate state is not changed during the autoionizing decay, $3s3d$ $^3\!D_3$ can not be populated when $j=3/2$.
Similarly, $3s3d$ $^3\!D_1$ can not be populated when $j=5/2$.
Consequently, at around 420 eV, only one strong resonance is expected for a) $^3\!D_1$ and c) $^3\!D_3$, whereas two resonances with an energy splitting of 0.5 eV are expected for b) $^3\!D_2$ and d) $^1\!D_2$.

The non-resonant excitation cross sections from $3s^k$ to $3s^{k-1}3d$ ($k=2$ for Fe$^{14+}$ and $k=1$ for Fe$^{15+}$) were also calculated including the cascading contribution from $3s^{k-1}4p$ and $3s^{k-1}4f$.

\section{\label{sec:results}Results and discussion}

\begin{figure}
\includegraphics[width=0.5\textwidth]{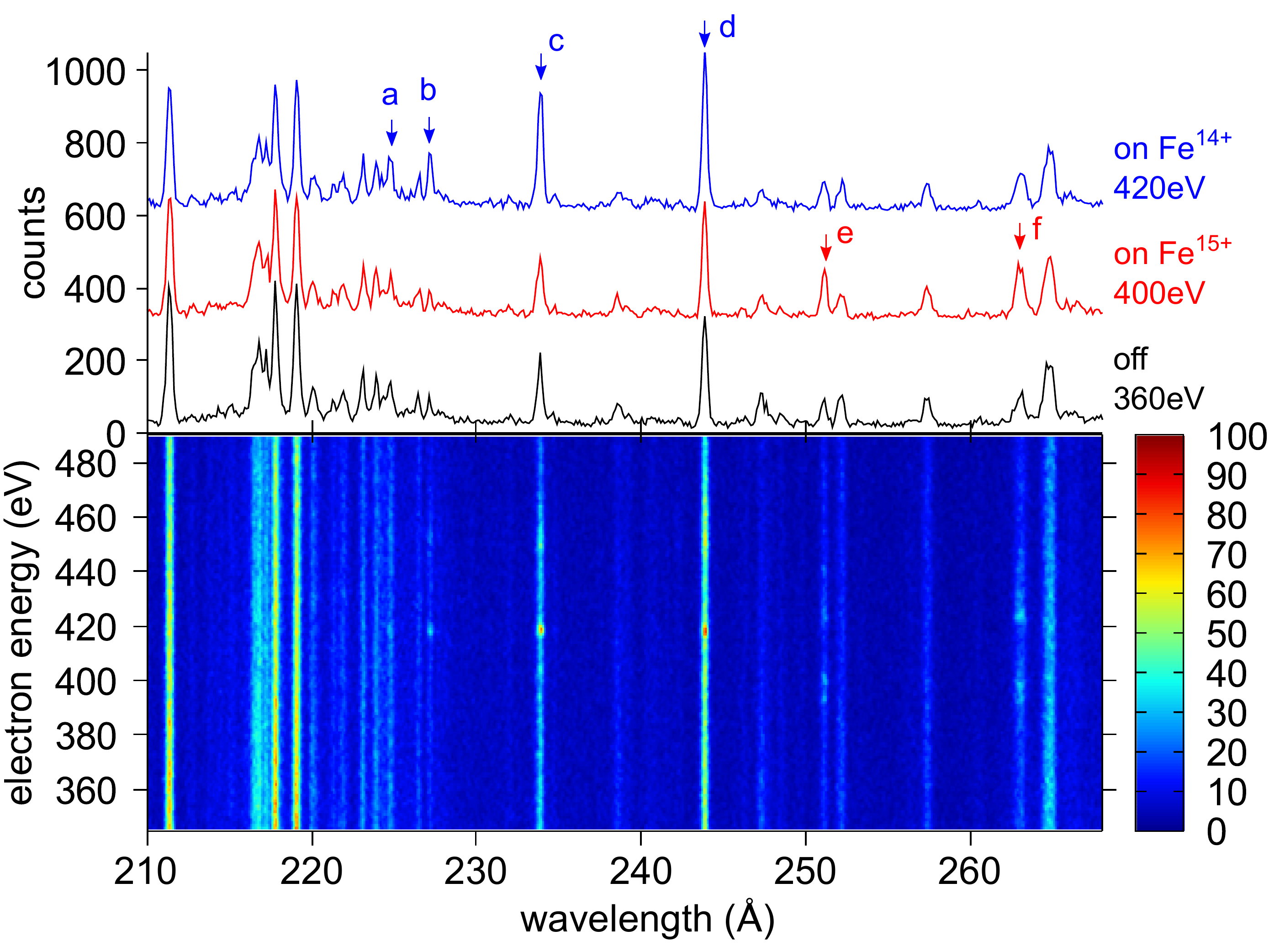}
\caption{\label{fig:2d}
Lower panel: two dimensional spectra showing the electron energy dependence of the observed EUV intensity.
Upper panel: EUV spectra sliced at electron energies of 360 (off-resonance energy), 400 (resonance energy for Fe$^{15+}$), and 420~eV (resonance energy for Fe$^{14+}$).
The labels a - f in the spectra correspond to the label in Table~\ref{tab:lines}.
}
\end{figure}

Figure~\ref{fig:2d} shows the energy dependent spectra obtained by a 45~h accumulation.
The absolute electron energy was normalized with the calculated resonance energy for Fe$^{14+}$ at around 420~eV.
As seen in the two dimensional map, enhancement of the EUV counts is found as bright spots at specific energies for the six lines labeled a to f.
The lines a-d are the $3s3p$--$3s3d$ transitions in Fe$^{14+}$ whereas the lines e and f are the $3p$--$3d$ transitions in Fe$^{15+}$ (see also Table~\ref{tab:lines}).
The enhancement of these lines can also be confirmed in the sliced spectra shown in the upper panel of Fig.~\ref{fig:2d}.

\begin{table}
\caption{\label{tab:lines}Lines of the present interest.
Wavelength values are taken from the CHIANTI database \citep{Dere1,Landi5}.
$R$ represents the calculated branching ratio in the radiative decay of the upper level.
}
\begin{ruledtabular}
\begin{tabular}{llcccc}
label&ion&lower&upper&wavelength (\AA)&$R$\\
\hline
a&14+&$3s3p$ $^3\!P_0$&$3s3d$ $^3\!D_1$&224.754&0.57\\
b&14+&$3s3p$ $^3\!P_1$&$3s3d$ $^3\!D_2$&227.208&0.77\\
c&14+&$3s3p$ $^3\!P_2$&$3s3d$ $^3\!D_3$&233.866&1.00\\
d&14+&$3s3p$ $^1\!P_1$&$3s3d$ $^1\!D_2$&243.794&0.99\\
e&15+&$3p_{1/2}$&$3d_{3/2}$&251.063&0.85\\
f&15+&$3p_{3/2}$&$3d_{5/2}$&262.976&1.00\\
\end{tabular}
\end{ruledtabular}
\end{table}

\subsection{Resonance in Fe$^{14+}$}

\begin{figure}
\includegraphics[width=0.5\textwidth]{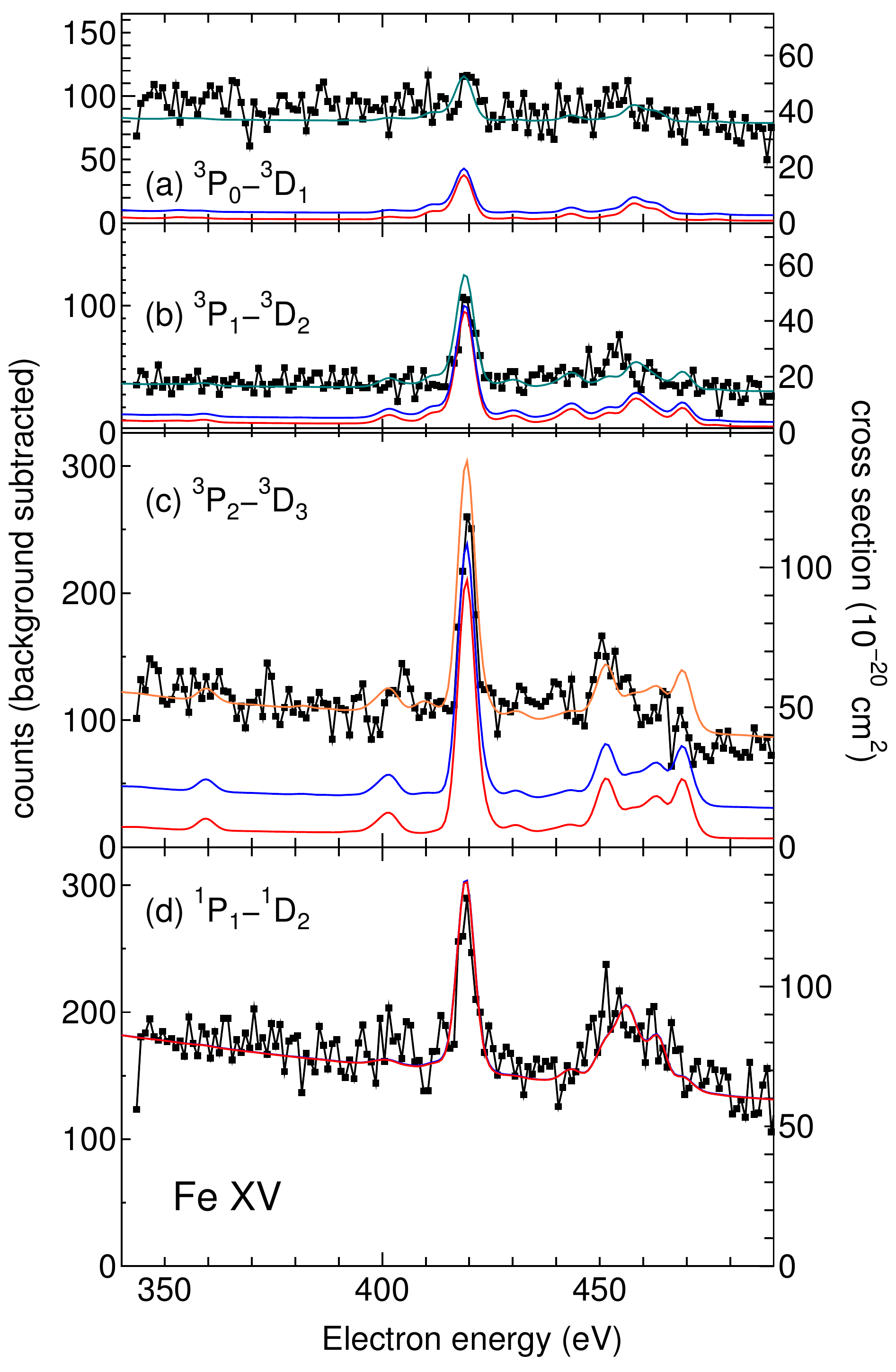}
\caption{\label{fig:XV}
Electron energy dependence of the experimental EUV photon counts for $3s3p$--$3s3d$ transitions in magnesiumlike Fe$^{14+}$ (black squares): (a) to (d) correspond to the label used in Fig.~\ref{fig:2d} and Table~\ref{tab:lines}.
Solid curves represent the calculated emission cross section obtained by assuming the population of $^3\!P_2$ to be nought (red), 4.8~\% (blue), and 16~\% (orange).
The green curves in (a) and (b) represent the emission cross section biased to fit the experimental intensity.
(see text for details of each curve.)
The left vertical axis is for the experimental counts, whereas the right vertical axis for the calculated cross sections.
}
\end{figure}

Figure~\ref{fig:XV} shows the experimental intensity of the four Fe$^{14+}$ lines plotted as a function of electron energy (black squares).
The background, mainly due to the dark counts of the MCP, was estimated from the wavelength region where no line was observed, and subtracted.
The correction on the relative quantum efficiency of the grating was made based on the data by \citet{Tu1}.
The correction was about 9 \% in the largest case between the lines (a) and (d).
The quantum efficiency of the MCP was assumed to be constant in this narrow wavelength range.

Since the collision frequency is much smaller than the radiative decay rate under the present electron density ($\sim10^{10}$~cm$^{-3}$), the experimental intensity is considered to be proportional to the emission cross section, which is the product of the excitation cross section ($\sigma^{\rm ex}$) for the upper level and the branching ratio $R$ for the transition of interest (The theoretical $R$ values are listed in Table~\ref{tab:lines}).
The calculated emission cross sections are also plotted in Fig.~\ref{fig:XV} as solid lines (right axis).
The red lines represent the cross sections obtained considering only the excitation from the $3s^2$ ground state.
Not only direct excitation to the $3s3d$ levels but also the cascading contributions via $3s4p$ and $3s4f$ levels are included, i.e. $\sigma^{\rm ex}=\sigma^{\rm ex}_{3s^2\rightarrow3s3d}+R_{4p\rightarrow3d}\cdot\sigma^{\rm ex}_{3s^2\rightarrow3s4p}+R_{4f\rightarrow3d}\cdot\sigma^{\rm ex}_{3s^2\rightarrow3s4f}$ is used for the red lines.
The blue lines represent the ``effective" cross sections obtained by taking also the excitation from the $3s3p$ levels into account, i.e. $n_{3s3p}\cdot\sigma^{\rm ex}_{3s3p\rightarrow3s3d}$ has been added to $\sigma^{\rm ex}$, where $n_{3s3p}$ represents the relative population of the $3s3p$ level with respect to the $3s^2$ ground state.
Among the four levels ($^3\!P_{0, 1, 2}$ and $^1\!P_1$) of the $3s3p$ configuration, $^3\!P_0$ and $^3\!P_2$ are metastable states that can have a non-negligible population even in a coronal plasma condition in CoBIT.
The population of these metastable states was estimated by a collisional radiative model (CRM) calculation to be 1.2~\% and 4.8~\% for $^3\!P_0$ and $^3\!P_2$, respectively for an electron energy of 600~eV (delta function distribution) and an electron density of $10^{10}$~cm$^{-3}$.
As seen in the figure, the inclusion of the metastable state contributions enhances the emission cross section especially for (c), for which most of the contribution arises from $^3\!P_2$.
For the other three lines, the metastable state contributions are insignificant.
It is noted that the blue line in (d) is hard to see due to the overlap with the red line.

According to the calculation shown in Fig.~\ref{fig:rs}, the resonance at around 420~eV in the $^3\!P_2$--$^3\!D_3$ transition (c) is considered to contain only one resonance via $[2p_{1/2}^{-1}3s^2 3p_{1/2}3d_{5/2}]_{J=5}$.
Thus the experimental width of this resonance is considered to represent the  energy width of the electron beam.
A full width at half maximum (FWHM) of 4.5 eV was obtained by fitting a Gaussian distribution to this resonance.
The calculated cross sections plotted in Fig.~\ref{fig:XV} were thus convoluted by a Gaussian distribution with a FWHM of 4.5 eV.
It is noted that the natural width of this resonance is too small (estimated to be $\sim 0.1$ eV) to contribute to the experimental energy width.

The right vertical axis was scaled so that the calculated cross section agrees with the experimental intensity at the non-resonant region (350 - 390~eV) for (d) $^1\!P_1$ -- $^1\!D_2$.
As seen in Fig.~\ref{fig:XV} (d), the experimental resonant features are well reproduced by the calculation under this scaling.
The same scaling is also applied for (a) to (c); in other words, the experimental intensity is normalized to the calculated non-resonant emission cross section of (d)  $^1\!P_1$ -- $^1\!D_2$.
As seen in the figure, under this normalization, the calculation obviously fails to reproduce the experimental intensity for the non-resonant region of (a) to (c).
The reason for this discrepancy may be the contamination from other lines.
Actually, the relatively weak lines (a) and (b) are not well separated from neighboring lines and seem to lie on a hump as confirmed in the spectra shown in the upper panel of Fig.~\ref{fig:2d}.
We thus do not go into detail about this discrepancy for the non-resonant region in (a) and (b).
The green curves in (a) and (b) represent the cross sections offset by a constant value to fit the experimental intensity for the comparison on the resonance strength (area of the resonance peak).
As seen in the figure, general good agreement is found between the experiment and the calculation for the resonance strength.

Unlike the lines (a) and (b), the prominent line (c) is well separated from other lines; thus the contamination is unlikely to occur.
The less possibility of the contamination to the line (c) was also confirmed in our previous study \citep{Shimizu2}.
Thus, from the present result, it can be concluded that the theoretical calculation underestimates the emission intensity (emission cross section) of (c) significantly in the non resonant region.
It can be caused by the underestimation of the excitation cross section for the upper level ($^3\!D_3$), the population of the lower level ($^3\!P_2$), or both.
The orange line shows the cross section assuming the population of the $^3\!P_2$ level to be 16~\%, i.e. more than three times as much as that predicted by the CRM calculation.
As seen in this example, a large modification on the excitation cross sections, the population, or both should be required to account for the discrepancy.
On the other hand, for the resonance strength (the area of the resonance peak) and resonance features, general agreement is found between the experiment and the calculation.

\begin{figure}
\includegraphics[width=0.45\textwidth]{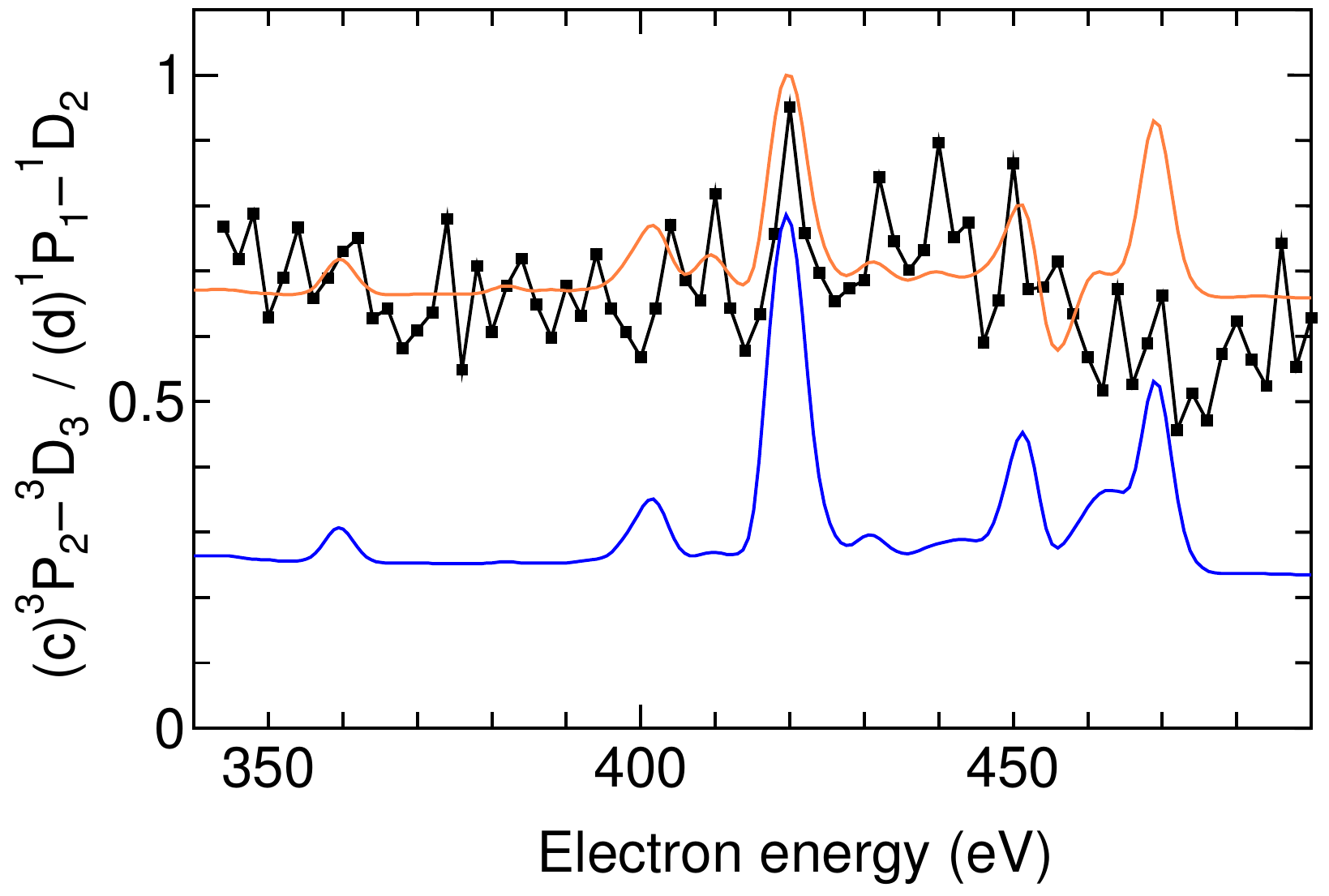}
\caption{\label{fig:ratio}
Electron energy dependence of the line ratio between (c) $^3\!P_2$ -- $^3\!D_3$ and (d) $^1\!P_1$ -- $^1\!D_2$.
Black squares represent the present measurement.
Solid lines are the theoretical emission cross section ratios assuming the population of the $^3\!P_2$ level to be 4.8~\% (blue) and 16~\% (orange).}
\end{figure}

In our previous studies \citep{Nakamura21,Shimizu2}, discrepancy between the experiment and model calculation was found for the intensity ratio of the line (c) to the line (d).
The previous experiment was done with an ``apparent" electron beam energy of 500 eV, which was determined just from the potential difference between the cathode and the DT2 electrode.
The actual electron beam energy could be lower by 20-30 eV due to the space charge potential of the electron beam; thus there was a suspicion that the RE contribution via $2p^53s^23d^2$ intermediate states could cause the discrepancy.
Figure ~\ref{fig:ratio} shows the ratio obtained from the present measurement as a function of electron energy (black squares).
The blue curve represents the theoretical emission cross section ratio (the ratio between the blue curves in Fig.~\ref{fig:XV} (c) and (d)).
The present result clearly shows that the contribution of the resonance to this ratio is insignificant, and that the discrepancy found in our previous study should be caused by the underestimation of the emission intensity of (c) in the non resonant region.
The orange curve represents the emission cross section ratio assuming the population of the $^3\!P_2$ level to be 16~\%, i.e. the ratio between the orange curve in Fig.~\ref{fig:XV} (c) and the red curve in (d) just for reference.

The present observation was made only at an observation angle of 90$^\circ$ with respect to the quasi-unidirectional electron beam.
Thus the anisotropic emission may cause the discrepancy.
We have estimated the angular distribution of the $^3\!P_2$--$^3\!D_3$ transition (c) using the flexible atomic code (FAC) \citep{FAC}.
The estimation showed that the deviation from the isotropic emission is as small as a few percent; thus it can not account for the discrepancy.

\subsection{Resonance in Fe$^{15+}$}

\begin{figure}[t]
\includegraphics[width=0.5\textwidth]{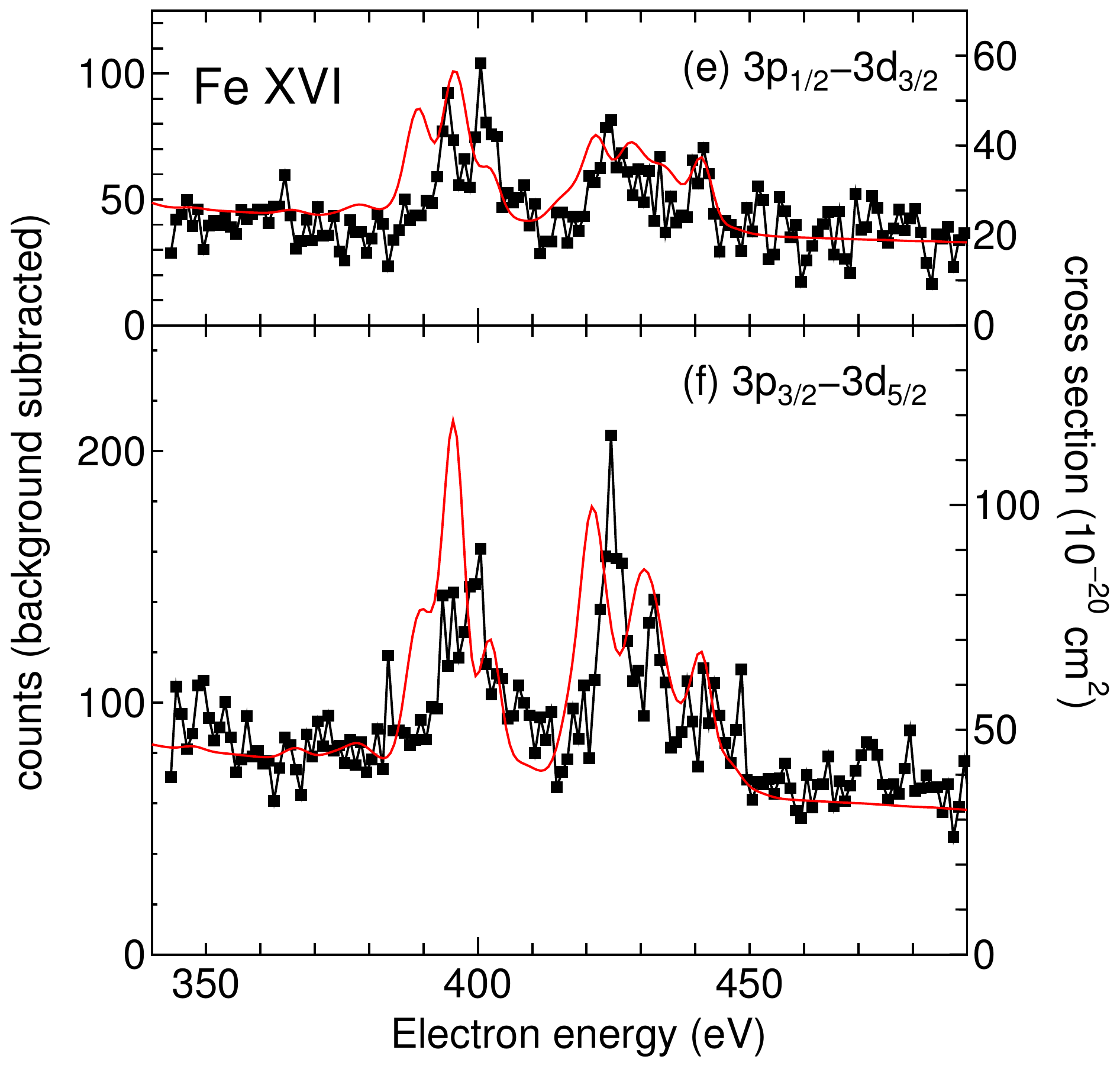}
\caption{\label{fig:XVI}
Electron energy dependence of the experimental EUV photon counts for $3p$--$3d$ transitions in sodiumlike Fe$^{15+}$ (black squares): (e) and (f) correspond to the label used in Fig.~\ref{fig:2d} and Table~\ref{tab:lines}.
Red solid curves represent the calculated emission cross section.
The left vertical axis is for the experimental counts, whereas the right vertical axis for the calculated cross sections.}
\end{figure}

Figure~\ref{fig:XVI} shows the experimental intensity of the two Fe$^{15+}$ lines as a function of electron energy (black squares).
Red curves represent the theoretical emission cross sections.
The features below 410~eV correspond to the resonances via $2p^53s3p3d$, whereas those above 410~eV correspond to the resonances via $2p^53s3d^2$.
Since there is no metastable state in the sodiumlike system, the excitation only from the ground state is considered.
The cascading contributions from $4p$ and $4f$ are included in the non-resonant cross sections.
The horizontal energy scale used in Fig.~\ref{fig:XV} is also applied to Fig.~\ref{fig:XVI}.
As seen in the figure, there seems to exist a 3-5~eV deviation in the peak position between the experiment and the calculation.
The experimental electron energy can not be determined absolutely, but the relative scale measured by a voltage meter is reliable.
It means that the calculation has an error of 3-5~eV for the relative resonance energy between Fe$^{14+}$ and Fe$^{15+}$. 
The vertical axis for the cross section (right axis) is scaled so that the experimental intensity agrees with the calculated emission cross section at the non-resonant region.
The same scaling is applied for both (e) and (f).
Unlike Fe$^{14+}$ (Fig.~\ref{fig:XV}), the agreement between the experimental intensity and the emission cross section curve is consistently good for both (e) and (f).
The agreement for the resonance features is also generally good except for the resonance at around 395~eV in (f), where the calculation seems to overestimate the experiment.
It corresponds to the resonance via the intermediate $[2p_{1/2}^{-1}3s3p_{1/2}3d_{5/2}]_{J=3}$ level.
We have calculated the magnetic sublevel distribution of this intermediate state using the FAC, and confirmed that the anisotropic distribution may enhance (not reduce) the emission at 90$^\circ$ by 17 \% compared to the isotropic distribution.
Thus the discrepancy in this resonance strength can not be explained by the anisotropy.
Further investigation is needed to account for this discrepancy.

\section{Summary}
We made laboratory measurements for resonant electron impact excitation of the $3d$ excited levels in Fe$^{14+}$ and Fe$^{15+}$ to provide a benchmark for the theoretical cross sections that are used to estimate the emissivity of the $3p$--$3d$ lines, which are important in the spectroscopic diagnostics of astrophysical plasmas.
The theoretical calculations based on the distorted wave approximation were confirmed to reproduce the experimental resonance features qualitatively.
On the other hand, a significant discrepancy was found in the non-resonant region of the emission cross section for the $3s3p$ $^3\!P_2$ -- $3s3d$ $^3\!D_3$ transition in Fe$^{14+}$.
We also studied the energy dependence of the intensity ratio between the $3s3p$ $^3\!P_2$ -- $3s3d$ $^3\!D_3$ and $3s3p$ $^1\!P_1$ -- $3s3d$ $^1\!D_2$ transitions, for which the inconsistency with the model has often been reported in observatory and laboratory measurements.
The present result showed that the contribution of the resonance to this ratio is insignificant.
The fundamental cause of the inconsistency in the intensity ratio is considered to be the underestimation of the theoretical emission intensity of the $3s3p$ $^3\!P_2$ -- $3s3d$ $^3\!D_3$ transition in the non-resonant region.
The atomic data (the excitation cross sections or the lifetime of the metastable $^3\!P_2$ or both) should be re-examined to solve the discrepancy.

\acknowledgments
This work was performed under the Research Cooperation Program in the National Institutes of Natural Sciences (NINS).

\bibliographystyle{yahapj}
\bibliography{ref}


\end{document}